# Functional parcellation using time courses of instantaneous connectivity

Erik S.B. van Oort[1], Maarten Mennes[1], Tobias Navarro Schröder[1,3], Vinod J. Kumar[4], Nestor I. Zaragoza Jimenez[1,5], Wolfgang Grodd[4], Christian F. Doeller[1,3], Christian F. Beckmann[1,2,6]

Affiliations

1 Donders Institute for Brain, Cognition and Behaviour, Radboud University, Nijmegen, The Netherlands

2 Radboud University Medical Centre, Department of Cognitive Neuroscience, Nijmegen, The Netherlands

3 Kavli Institute for Systems Neuroscience and Centre for the Biology of Memory, Norwegian University of Science and Technology, NTNU, 7491 Trondheim, Norway;

4 Max Planck Institute for Biological Cybernetics, Tübingen, Germany

5 Max Planck Institute for Human Cognitive and Brain Sciences, Department of Neuropsychology, Leipzig , Germany

6 Oxford Centre for Functional Magnetic Resonance Imaging of the Brain (FMRIB), University of Oxford, Oxford, OX3 9DU, United Kingdom.

# Abstract

Functional neuroimaging studies have lead to understanding the brain as a collection of spatially segregated functional networks. It is thought that each of these networks is in turn composed of a set of distinct sub-regions that together support each network's function. Considering the sub-regions to be an essential part of the brain's functional architecture, several strategies have been put forward that aim at identifying the functional sub-units of the brain by means of functional parcellations. Current parcellation strategies typically employ a bottom-up strategy, creating a parcellation by clustering smaller units. We propose a novel top-down parcellation strategy, using time courses of instantaneous connectivity to subdivide an initial region of interest into sub-regions. We use split-half reproducibility to choose the optimal number of sub-regions.

We apply our Instantaneous Connectivity Parcellation (ICP) strategy on high-quality resting-state FMRI data, and demonstrate the ability to generate parcellations for thalamus, entorhinal cortex, motor cortex, and subcortex including brainstem and striatum. We evaluate the subdivisions against available cytoarchitecture maps to show that the our parcellation strategy recovers biologically valid subdivisions that adhere to known cytoarchitectural features.

## 1. Introduction

Macroscopically observable brain function is hypothesized to rely on interactions within and between hierarchically organized sets of brain regions (Bellec et al., 2006; Sporns, 2011). Within this framework, smaller, functionally specialised units interact in the context of large-scale networks that connect distant regions within the brain (Sporns, 2011). Such subunits are believed to follow topographic organisational principles with representations at multiple scales within a full functional hierarchy (Smith et al., 2012).

Support for this view on brain function comes from critical advances in imaging neuroscience that exploit structural and functional connectivity analyses. In order to enable inferences about the intrinsic organisation of a functional network, connectivity analyses typically proceed by defining a selection of regions of interest (ROIs) hypothesized to represent key area within the larger network under investigation. Understanding the structural and functional properties of these ROIs is key to describe total network function, yet the selection of ROIs can greatly influence the accuracy of the inferred network organisation (Smith et al., 2011). Inaccurate ROI location or outlines will lead to a mixing of signal from different sub-regions, yielding inaccurate characterization of the underlying functional biology. Here we introduce a novel, principled method for top-down in-vivo functional brain parcellation and demonstrate that network organisation can be characterised in terms of a hierarchy of functional sub-regions. Further, we demonstrate that these sub-regions closely correspond to known subdivisions (e.g. on the basis of cytoarchitectonics) across a number of model systems.

Current descriptions of the brain's functional architecture typically start with the identification of a large number of (small) functional units, regions or atoms (de Reus & van den Heuvel, 2013). Subsequently, these regions are grouped together into larger networks, e.g., by means of employing clustering approaches. As grouping can occur at multiple levels, the initially defined units will be combined into a multi-level functional hierarchy. The idea of grouping smaller units into larger networks is commonly referred to as a *bottom-up* parcellation strategy and has been successfully demonstrated (e.g. on the basis of random parcellation (Hagmann et al., 2008), *k*-means clustering (Craddock, James, Holtzheimer, Hu, & Mayberg, 2012; M. van den Heuvel, Mandl, & Hulshoff Pol, 2008), region growing (Blumensath et al., 2013), or high-model order independent component analysis (ICA) (Beckmann, 2012; Beckmann & Smith, 2004; Kiviniemi et al., 2009). Parcellations from such bottom-up strategies are fully dependent on the initial definition of the smallest functional entities. Ideally, these smallest entities directly relate to valid neurobiological quantities, such as functionally defined areas or cytoarchitectonically homogeneous patches of cortex. In the absence of biological validity, any inaccuracies or ambiguities in the initial parcellation can propagate through the ensuing hierarchical characterisation of the brain's functional organisation. Defining such elementary building blocks is faced with substantial challenges. First, cortical areas of homogeneous function are believed to vary in size by several orders of magnitude (David C. Van Essen, Glasser, Dierker, Harwell, & Coalson, 2012). Any parcellation strategy that gravitates towards areas of uniform size such as 'random parcellations' will therefore inherently be inadequate to accurately model the underlying biology. Second, while several areas of the primate brain are clearly separable from adjacent areas in terms of their functional properties, other adjacent areas show more subtle differences (David C. Van Essen et al., 2012). Both overlapping organisations (e.g. within the visual system) and gradual transitions between functionally homogeneous patches (e.g. within multimodal association cortices) remain a modelling challenge for classic parcellation strategies and limit the granularity of any population-level result. Finally, as a result of inter-subject variability in the brain's functional and structural architecture, any ensuing registration inaccuracies will impose limitations on the size of the smallest subunits that can robustly be detected at the population level, even in light of advanced techniques for aligning multiple participants' brains (Robinson et al., 2014). Jointly these challenges complicate the question of size, locality, topography and number of identifiable areas in existing brain parcellations (Thirion et al., 2006).

In this work, we demonstrate a principled approach to define functional parcellations in a *top-down* manner called *Instantaneous Connectivity Parcellation*, or ICP. With ICP, we transform data within a ROI to make it more sensitive to spatial differences in temporal dynamics within that ROI. To illustrate our approach we applied ICP to resting state data available in the Human Connectome Project and 7T FMRI data of a spatial navigation task available locally at the Donders Institute. We derived parcellations for several key brain areas, i.e., sensory-motor cortex, subcortex (brainstem and striatum), thalamus, and entorhinal cortex. To assess the biological validity of the obtained parcellations we compare the obtained regions with histological atlases, a currently available parcellation (Craddock et al., 2012), and parcellations obtained by ICA without prior temporal unfolding using Dice's overlap (Dice, 1945).

## 2. Material and Methods

### 2.1. Data

To illustrate and evaluate ICP we used data from 100 subjects of the Human Connectome Project (D. C. Van Essen et al., 2012) as well as data acquired at 7T (Navarro Schröder, Haak, Zaragoza Jimenez, Beckmann, & Doeller, 2015). The publically available Human Connectome Project data consists of a very extensive list of MRI modalities and behavioural scores. The resting state FMRI (R-FMRI) data consists of two sessions, with two scans of fifteen minutes each, for a total of one hour scanning time for each subject. A multiband protocol with a very rapid TR of 720ms resulted in a total of 4800 time points for each subject. This provides a large number of temporal degrees of freedom, making it ideal for the exploitation of R-FMRI dynamics. Data quality was further improved using a new automated ICA-based denoising technique called FSL-FIX, removing a wide range of artefacts (Griffanti et al., 2014; Salimi-Khorshidi et al., 2014). HCP subject ID's that were included in our study are available in the supplementary material. For a description of the 7T dataset that was used to obtain a parcellation of entorhinal cortex we refer to Navarro Schroder et al 2015.

### 2.2. Instantaneous Connectivity Parcellation

Our proposed parcellation strategy consists of multiple steps. First, to provide a robust basis for the parcellation we start from an initial large-scale ROI that is well grounded on prior anatomical or functional knowledge, and ideally is selected in light of maximal population-level reproducibility and biological validity (e.g., a resting state network, anatomical ROI, region determined by functional localizer, whole-brain mask, etc.). We aim to divide the large-scale ROI into smaller, functionally homogenous sub-regions based on their temporal signature. In order to define such sub-regions we need to identify sub-region-dependent changes in the connectivity profile. In order to make the analysis sensitive to such changes, we analyse the dynamics of the 'instantaneous' modes of connectivity, reflecting the voxel-to-region differences in functional connectivity. In the case of time series correlations as a primary measure of 'connectivity', this amounts to transforming the FMRI time series into instantaneous correlation values, though this approach can be used with other definitions of connectivity as soon as these are time-averaged values.

Basic Pearson correlations are most widely used to quantify degrees of connectivity and have seen extensive use both in defining RSN maps, and estimating functional connectivity between ROIs (Biswal, Yetkin, Haughton, & Hyde, 1995; Cole, Smith, & Beckmann, 2010). In such types of analysis the ensuing quantities are based on temporal averages. This averaging hides the rich dynamic information present in resting BOLD data. With our ICP strategy we expand upon the basic Pearson correlation by considering the sequence of events across time, which ultimately averaged across the length of the experiment result in a time-averaged correlation map. The Pearson correlation between time courses $x$ and $y$ is defined as

$$\rho_{x,y} = \frac{E[(x(n) - \mu_x)(y(n) - \mu_y)]}{\sigma_x \sigma_y}$$

with μ the mean and σ the standard deviation of each time course. In this example $x$ can be the time course of a seed voxel, and $y$ the time course of a target voxel. For normalised time series (mean zero and unit standard deviation) this changes to

$$\rho_{x,y} = \frac{1}{T} \sum_{n=1}^{T} x(n) y(n)$$

where T is equal to the number of time points in the measurement. In other words, the Pearson correlation is equal to the (temporal) mean of the element-wise (or Hadamard) product between two normalised time courses, i.e.

$$\begin{pmatrix} \rho_{x,y}(1) = x(1)y(1) \\ \rho_{x,y}(2) = x(2)y(2) \\ \rho_{x,y}(3) = x(3)y(3) \\ \vdots \\ \rho_{x,y}(T) = x(T)y(T) \end{pmatrix}$$

can be considered as the time series of instantaneous connectivity between the reference time course $x$ and target time course $y$. This vector is the result of temporal unfolding of the Pearson correlation and contains all the dynamic information that on average results in the static correlation value.

With the ICP method we use the average time course of the original region selected for parcellation as the reference time course $x$ and analyse the time-resolved instantaneous connectivity between this regionally-specific reference time series and all voxels' time series within the same region. For regions that at a coarse scale appear functionally homogeneous a simple seed-based correlation map would simply recover the selected region. This implies that the dynamic information depicted by the instantaneous correlations between this reference and all voxels of the selected region on average reflects that aspect of the data, which contributes to the spatial delineation of this region relative to other coarsely defined areas (possible starting ROIs). The instantaneous connectivity transformation of all voxels' data on the basis of this one reference time series *x* therefore acts as a filter, emphasizing subtle differences relative to the average dynamics within this specific region, and deemphasizing outside influences. To illustrate this effect consider the example in figure 1: the example shows a simplified situation where there is a single region of interest (black outline) with four distinct substructures that are yet to be identified. These sub-structures necessarily have sub-region specific temporal dynamics that are nearly identical, with transient events at different time points. We assume that each of the four substructures show spatially structured temporal events, represented in this example by transient 'events' (here, the different sub-regions show small, yet identifiable peaks). A simple correlation analysis cannot distinguish between the time courses, as they have near identical correlation coefficients (in the case of this example: 0.88) between them. The correlations between the average time course and the four region-specific time courses are also identical (0.954). Transforming the original data by means of temporal unfolding to the instantaneous connectivity values, it is immediately apparent that the transient events of the original time series become amplified and time periods where a specific sub-region is effectively driving the overall ROI reference time course feature more strongly. This is also reflected in the SNR of these events, which goes up from 1.06 in the original time series to 1.18 (11% increase) in the instantaneous connectivity time series. Assuming these transient events do appear in a spatially structured fashion (i.e. give rise to sub-regions within the ROI), we can use a multivariate method like ICA to delineate the corresponding spatial maps.

An alternative approach to temporal unfolding would be to effectively remove the average ROI time course by means of time-series regression (see supplementary material). While this would also emphasize the sub-region specific periods in the data, the resulting transformed time series data would have reduced variance in cases where transient events are sparse.

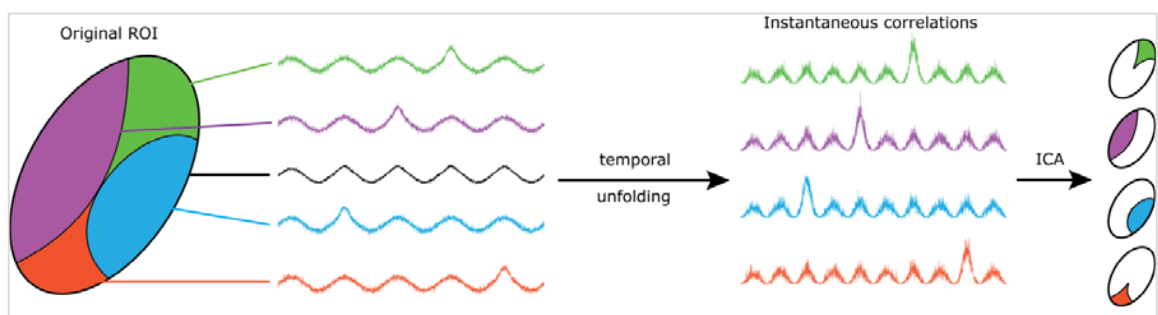

```
Figure 1, Simulated time courses for an example ROI and its potential subdivisions.
The time courses were made using a simple sinusoid, a transient event, and additive
Gaussian noise. Each time course represents a voxel from a subdivision within the
larger ROI. The transient events are spatially structured across these subdivisions.
A Pearson correlation analysis cannot distinguish between them, as they have the
exact same correlation coefficient with each other and with the mean time course.
The instantaneous correlation time courses are shown on the right. It is clear that
the temporal unfolding increases the SNR of the transient events making them more
detectable for subsequent grouping e.g., using ICA.
```

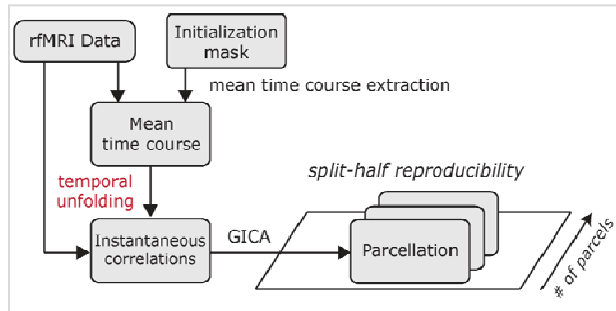

**Figure 2**, ICP pipeline: first, each individual participant's data is combined with
the ROI initialization mask in order to obtain the mean time course within the ROI.
Both the voxel-wise data and the mean time course are normalized and combined
(element-wise multiplication) to generate temporally unfolded measures of
instantaneous connectivity for each voxel and each participant. Each participant's
transformed data is fed into a group ICA to generate a parcellation of the initial
ROI. The group ICA is performed multiple times across random splits of participants
and increasing dimensionality (i.e., the number of returned components) in order to
determine the parcellation that is most reproducible across participants. The
complete analysis pipeline is fully data-driven, and requires little user input,
only preprocessed R-FMRI data along with a region of interest mask needs to be
supplied.

### 2.3. Split-half reproducibility

A strategy is required to decide which parcellation is best supported by the data and the underlying functional organisation of the network of interest. We address this issue by using split-half reproducibility to decide on the scale (i.e., number of parcels) that is most reproducible given the data, and use this scale for the final parcellation. The final group ICA step is repeated across random splits into groups of subjects across a range of scales. The split-half reproducibility at each scale is calculated by way of Dice's overlap between the resulting parcellations of each split. As these components are not necessarily ordered identically, they are reordered in order to find the best matching pair across the split. A strong on- and weak off-diagonal in the resulting overlap matrix indicates a strong one-to-one match between the parcellations of each split. The reliability at each scale was estimated by taking the average of the on-diagonal of this Dice overlap matrix. This process was iterated multiple times (20 iterations in this work) to get reliable averages at each scale. Figure 2 illustrates the full ICP pipeline including the split-half reproducibility.

### 2.4. Regions of interest

We demonstrate the potential of ICP by creating parcellations of sensory-motor cortex and subcortex including brainstem and striatum. Further, to evaluate the ability to delineate functional sub-regions within small and confined ROIs we illustrate the ability to parcellate thalamus and entorhinal cortex. For the sensory-motor system, we defined the initial region of interest by first applying ICP to a full brain mask, resulting in nine large-scale parcellations that corresponded to well known resting state networks. Supplementary Figure 2 provides an overview of these large-scale parcels. Subsequently we applied ICP to the parcellation that represented the sensory-motor system, including both pre- and post-central gyrus, as well as supplementary motor area (SMA) across both hemispheres. For the subcortex parcellation we based our ROI on the Harvard-Oxford atlas (Desikan et al., 2006; Frazier et al., 2005; Goldstein et al., 2007; Makris et al., 2006). For the thalamus parcellation applied ICP to a bilateral anatomical mask of the thalamus as defined by the Morel histological atlas (Morel, Magnin, & Jeanmonod, 1997). Similarly, we used a manually delineated anatomical starting mask (Navarro Schröder et al., 2015) for the parcellation of entorhinal cortex.

### 2.5. Evaluating resulting parcels

To evaluate the validity of the resulting parcels we compared our results against already available information in the form of cytoarchitectonic atlases (where available). The parcels obtained in the sensory-motor system were compared against the Jülich histological atlas (Amunts et al., 1999) as provided by FSL (FMRIB Software Library, (Jenkinson, Beckmann, Behrens, Woolrich, & Smith, 2012; Smith et al., 2004; Woolrich et al., 2009)). Similarly we evaluated the thalamus parcellation against subdivisions available in the Morel atlas that was used to derive the initial ROI for the Thalamus (Morel et al., 1997). For the entorhinal cortex parcellation we compared the ensuing results qualitatively to a 2D visual diagram of entorhinal organisation given by Krimer et al. (1991). Finally, for the subcortical regions we had no cytoarchitectonic atlas available, instead we compared our results against the sub-regions as available in the Harvard-Oxford atlas provided by FSL.

To evaluate how IC-parcellations compared to alternative parcellation strategies we also compared an existing parcellation obtained using NCUT clustering (Craddock et al., 2012) to the atlases. For each initial structure we used the NCUT parcellation that yielded a similar number of clusters as obtained using our ICP parcellation in our region of interest. Similarly, we compared the available atlases against parcels obtained by applying ICA to data without temporal unfolding. For each initial structure we obtained an ICA parcellation with a dimensionality that matched the ICP scale of choice.

All comparisons were conducted by means of calculating Dice's overlap between atlas regions and the closest matching parcels (Dice, 1945). Dice's overlap is calculated as twice the intersection between regions divided by the union of both regions.

## 3. Results

### 3.1. Sensory-motor system

We obtained a 12-sub-region functional parcellation of the sensory-motor system (see figure 3). The overall structure of the parcellation showed a high degree of left-right symmetry, and distinguished primary somato-sensory cortex, primary motor cortex, and pre-motor cortex.

Our parcellation recovered two major principles of motor cortex organisation. First, sensory-motor cortex including anatomically defined (e.g. by the Jülich atlas (Amunts et al., 1999)) sub-regions (e.g., primary motor cortex) which are typically arranged along the anterior-posterior axis. Figure 3 illustrates this for our parcellation by direct comparison with the Julich histological atlas. This comparison was achieved by projecting both our parcellation as well as the atlas to a flat cortical surface using the Caret software package (David C. Van Essen et al., 2001). Next to the anterior-posterior organisation, the sensory-motor system can also be organized according to function. This organisation is typically illustrated by means of a homunculus referring to the relative mapping of the human sensory-motor system (hands, feet, lips…) to areas in the brain's sensory motor cortex (Penfield & Boldrey, 1937). This organization lies along the central sulcus from the medial to the lateral wall and can also be observed in our parcellation solution.

Table 1 lists Dice's overlap scores between the Jülich atlas and selected parcels obtained from the ICP and ICA-based parcellations, as well as through NCUT clustering. We chose scale 50 from the NCUT parcellation of Craddock et al. (2012) as it most closely matched the 12 parcels within the motor system. The comparison with the atlas shows that the IC parcellation was able to retrieve several regions that exhibited a good match in shape, size and location between the atlas and the parcels, especially for parcels that corresponded to the medial regions of the motor system, including primary motor cortex, premotor cortex superior parietal lobule and secondary somatosensory cortex. These regions are all relatively confined along the primary motor strip. In contrast, for the lateral elements of the motor system where the organization is focussed on the split across the central sulcus, the IC parcellation and Jülich atlas start to differ. Interestingly, this is where the functional parcellation more closely follows the functional organization of the homunculus. Compared to the IC parcellation, the ICA parcellation resulted in similar overlap scores, while the Craddock parcellation did not match well with the Jülich atlas regions.

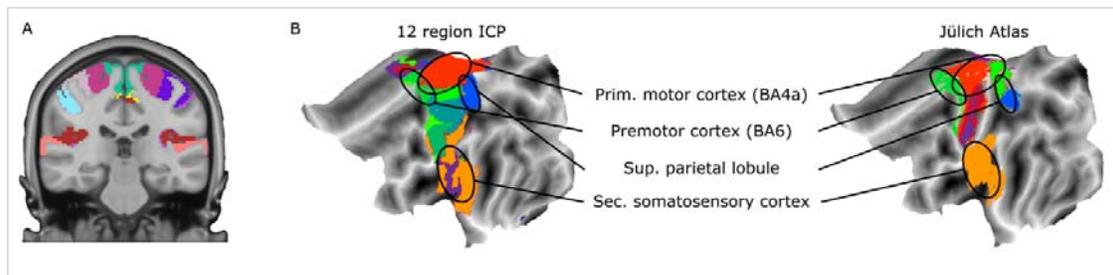

**Figure 3**: IC-Parcellation of the Sensory-Motor system that yielded 12 sub-regions. **A.** Coronal slice (y=-28) that displays the obtained sub-regions. **B.** The flattened

cortical surfaces illustrate the correspondence between the IC-parcellation and the Jülich atlas. Several regions closely matched between parcellation and atlas. Yet, the IC parcellation also adhered to a medial-lateral organization that resembled the functional organization of the human homunculus.

| Anatomical region | ICP Dice | ICA Dice | Craddock scale 50 Dice |
|---|---|---|---|
| Secondary somatosensory cortex / Parietal operculum | 0.9 | 0.9 | 0.75 |
| Superior parietal lobule 5L | 0.68 | 0.68 | 0.53 |
| Premotor cortex BA6 | 0.67 | 0.62 | 0.44 |
| Primary motor cortex BA4a | 0.54 | 0.61 | 0.60 |
| Primary motor cortex BA4p | 0.49 | 0.42 | 0.3 |
| Superior parietal lobule 7A | 0.35 | 0.31 | 0.56 |
| Primary somatosensory cortex BA3a | 0.3 | 0.33 | 0 |

**Table 1**: Dice overlap between Jülich atlas and Sensory-Motor system parcellation, ICA and Craddock parcellation.

### 3.2. Subcortical structures

The subcortical structures of the brain include several key areas of the human connectome. Yet, their size and location in the brain make them a challenging target to investigate their potential parcellation structure. To further demonstrate the utility of ICP, we parcellated a subcortical ROI that included brainstem as well as the striatal regions.

Using ICP we obtained two different parcellations (figure 4), as examination of the split-half reproducibility revealed a clear maximum at scale 8 and a more local maximum at scale 27 (see supplementary figure 4). Figure 4 illustrates both parcellation solutions where the difference between the lower and the higher scale parcellation can be interpreted in a hierarchical manner: at the lower scale larger, structures such as the hippocampal formation are observed as single parcel, while at the higher scale these areas break down into biologically meaningful sub-divisions, with e.g., the hippocampal formation being parcellated into the hippocampus proper and amygdala. In the case of the brainstem the higher order parcellations resulted in a granular subdivision with high left-right symmetry. Especially in the brainstem the mesencephalon, anterior and posterior parts of the pons and the cerebellar peduncles can now be differentiated.

The Dice overlap between the IC-parcellation containing 27 sub-regions and the Harvard-Oxford atlas representation of the brainstem and sub-cortical regions is shown in table 2. This table also shows the overlap between the atlas regions and the parcellation of Craddock et al. (2012) at scale 50, as well as with a parcellation obtained using ICA with a dimensionality of 27 on data that were not temporally unfolded (as in ICP). However, overlap calculations are potentially adversely influenced when there is a large degree of difference in the granularity between the parcellation and its corresponding atlas region. Even at scale 8, we observed distinct subdivisions in the brainstem, corresponding to midbrain, medulla and pons, where the Harvard-Oxford atlas only contains a single brainstem region. This mismatch is even greater at scale 27, where the parcellation shows multiple nuclei-like regions in the brainstem. For this reason, we combined parcels for regions such as brainstem, in order to match the higher degree of granularity of the ICP and ICA parcellation to the coarser granularity of the atlas region. In the case of brainstem, parcels overlapping brainstem were combined until an optimal fit was found. Table 2 illustrates how ICP reaches a 24% higher degree of overlap (average Dice 0.67) when compared to ICA (average Dice 0.54), and a 131% higher degree of overlap with the atlas regions compared to the Craddock parcellation (average Dice 0.29).

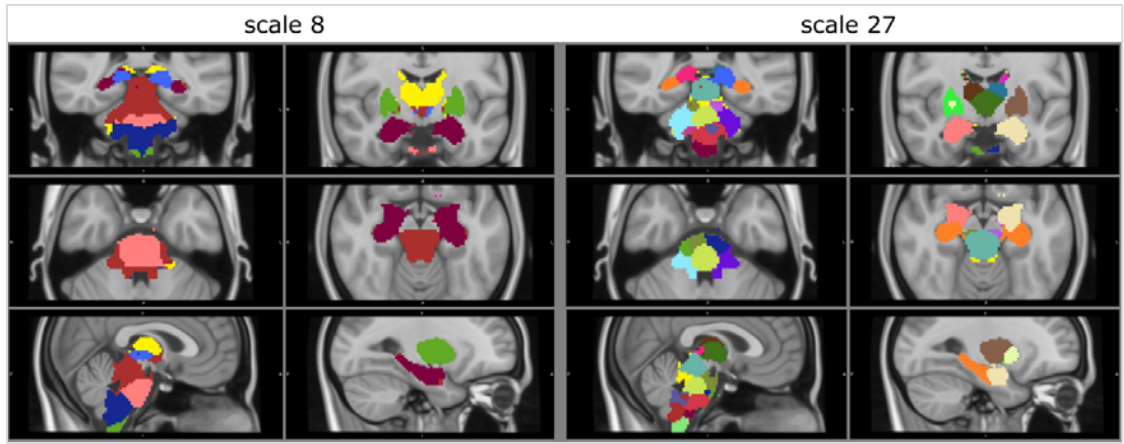

**Figure 4:** IC-parcellation of the subcortical regions at scale 8 and scale 27. The lower scale parcellation (scale 8) shows recognizable larger structures, including the thalamus, hippocampal formation, midbrain, medulla, and pons. For the higher scale parcellation (scale 27) these larger structures break down into their smaller sub-components (e.g. hippocampal formation into hippocampus proper and amygdala).

| Anatomical region | ICP Dice | ICA Dice | Craddock scale 50 Dice |
|---|---|---|---|
| Brainstem | 0.93 | 0.86 | 0.76 |
| Thalamus | 0.91 | 0.58 | 0.75 |
| Caudate | 0.82 | 0.62 | 0.17 |
| Putamen | 0.74 | 0.74 | 0.29 |
| Amygdala | 0.66 | 0.47 | 0.12 |
| Hippocampus | 0.63 | 0.63 | 0.26 |
| Pallidum | 0.47 | 0.27 | 0 |
| Nucleus Accumbens | 0.22 | 0.17 | 0 |

**Table 2:** Dice's overlap between Harvard-Oxford atlas and parcellations based on ICP and ICA. The IC-parcellation shows a higher degree of overlap than the ICA based parcellation.

### 3.3. Thalamus

Due to its projections to the entire cortex, the human thalamus is an essential node in any analysis to understand the brain as a complex network of interacting nodes. It is known from in- and ex-vivo studies that the thalamus contains several subdivisions and internal nuclei, each substructure exhibiting different connectivity profiles to the rest of cortex (Morel et al., 1997).

We applied ICP to a bilateral anatomical mask of the thalamus as defined by the Morel histological atlas (Morel et al., 1997), obtaining a parcellation containing 30 sub-regions (Figure 5). Validating this parcellation against the 32 cytoarchitectonic sub-regions delineated within the Morel atlas revealed large overlap (Figure 5). We confirmed the visual similarity by calculating Dice's overlap between the obtained parcellation and the regions defined by the atlas (Table 3). The pulvinar, lateral geniculate (LGN), sub-thalamic, and red nuclei exhibit excellent overlap above 0.74. Note that while they are not considered to be part of thalamus proper, the sub-thalamic and red nucleus are part of the Morel atlas, and are therefore also contained in the parcellation. In contrast, overlap between ICP and the Morel atlas was lower in the internal lamina structure, which is part of the central-lateral nucleus. While the internal lamina structure is regarded as a homogenous structure in the Morel atlas, ICP did not fully differentiate this structure from its surrounding nuclei. For further comparison, the Morel atlas was also compared with a parcellation based on masked ICA of the same data (Table 3). When ICA and ICP are compared it becomes clear that ICP shows a 51% higher degree of overlap (average Dice 0.59) when compared to ICA (average Dice 0.39) for all but one of the Morel regions. The difference with the Craddock parcellation at scale 900 is even larger, with a 73% higher degree of overlap for ICP compared to the Craddock regions (average Dice 0.34).

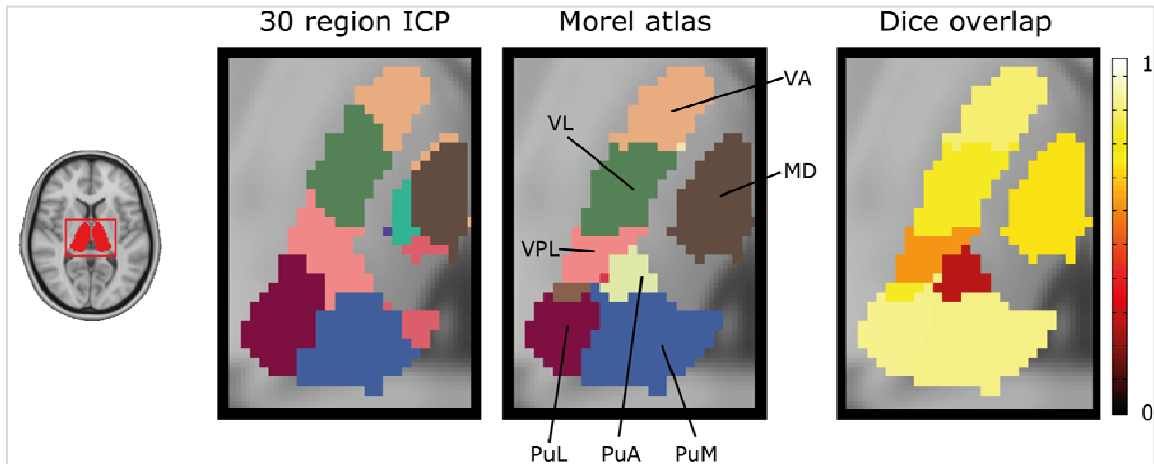

**Figure 5:** IC-Parcellation of human thalamus (left) compared to Morel atlas (middle). Several atlas nuclei are labelled. These are: Lateral Pulvinar (PuL), Medial Pulvinar (PuM), Ventral Anterior (VA), Ventral Lateral (VL), Medial Dorsal (MD), Ventral Posterior Lateral (VPL), and Anterior Pulvinar (PuA). The internal lamina structure was masked out, as this is a predominantly white-matter structure. Dice overlap between the IC-Parcellation and Morel atlas is shown on the right.

| Nuclei | ICP Dice | ICA Dice | Craddock parcellation scale 900 Dice |
|---|---|---|---|
| Red Nucleus | 0.96 | 0.69 | 0.34 |
| Sub-Thalamic | 0.91 | 0.6 | 0.11 |
| Lateral Geniculate | 0.87 | 0.34 | 0.18 |
| Lateral Pulvinar | 0.76 | 0.53 | 0.31 |
| Medial Pulvinar | 0.74 | 0.54 | 0.55 |
| Ventral Anterior | 0.69 | 0.42 | 0.65 |
| Lateral Posterior | 0.65 | 0.35 | 0.31 |
| Ventral Lateral | 0.64 | 0.54 | 0.61 |
| Medial Dorsal | 0.61 | 0.64 | 0.32 |
| Central Lateral | 0.54 | 0.28 | 0.48 |
| Medial Geniculate | 0.44 | 0.44 | 0.51 |
| Ventral Posterior Lateral | 0.44 | 0.17 | 0.26 |
| Parafasicular | 0.28 | 0.22 | 0.23 |
| Central Medial | 0.25 | 0.17 | 0.37 |
| Anterior Pulvinar | 0.18 | 0.17 | 0.09 |
| Ventral Posterior Inferior | 0.12 | 0.11 | 0.10 |

**Table 3:** Dice's overlap ICP parcellation, ICA and the Craddock parcellation at scale 900 compared against the selected nuclei of the Morel atlas.

## Entorhinal cortex

The Entorhinal Cortex (EC) is commonly perceived as a major input and output structure of the hippocampal formation, functioning as a node within cortico-hippocampal circuits (Canto, Wouterlood, & Witter, 2008). Cytoarchitectonically, up to nine subdivisions of the human EC have been distinguished (Krimer, Hyde, Herman, & Saunders, 1997). Starting from a manually defined anatomical mask, we created a parcellation of EC using high resolution FMRI data (0.9mm isotropic voxels) acquired at 7 Tesla in participants performing a spatial navigation task. (Navarro Schröder et al., 2015).

Figure 6 illustrates the EC parcellation obtained with ICP. Similar to what we observed for thalamus, the ensuing parcellation corresponds well with the size, location and shape of known cytoarchitectonic subdivisions (Krimer et al., 1997). Note that the parcellation is not identical to available images from cytoarchitectonics. Notable differences are the location of the prorhinal area (Pr) and the apparent absence of the intermediate-caudal area (Ic) in our parcellation. These differences might be related to inter-individual variability, as the cytoarchitectonic 2D drawing was derived post-mortem based on eight participants, whereas the ICP parcellation is based on in-vivo FMRI data from 22 participants. As for the other systems discussed in this work we aim to compare parcellation results against known underlying histology. However, because there is to our knowledge no established histological atlas for the EC, we restrict ourselves to a qualitative comparison against the work by Krimer et al. (1997). Notably, despite a wealth of data on the function of EC in animals, little is known about the function of the sub-regions in EC in humans where investigations are mainly complicated by cortical folding patterns that obscure how animal homologue areas map to the human brain. The ability of ICP to obtain biologically plausible results in this difficult to map area highlights the potential of our parcellation strategy.

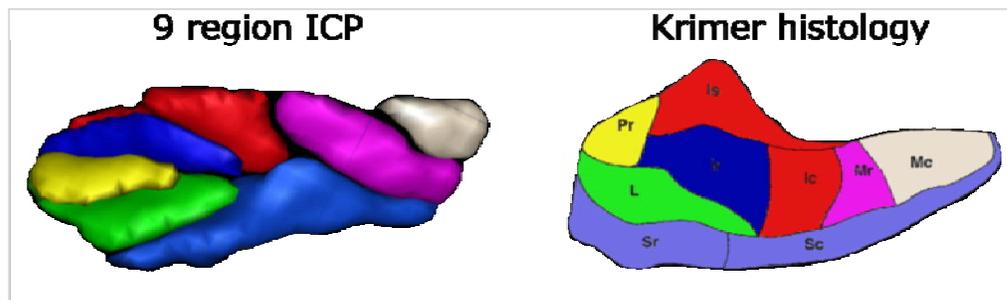

**Figure 6:** Volumetric representation of the IC-Parcellation of the human entorhinal cortex (left) compared to known cytoarchitectonic organization (right) based on 2D drawing from Krimer et al. (1997). Shown here are the Prorhinal (Pr), Lateral (L), Sulcal rostral (Sr), Sulcal central (Sc), Intermediate superior (Is), Intermediate rostral (Ir), Intermediate central (Ic), Medial rostral (Mr) and Medial caudal (Mc) regions of entorhinal cortex.

## 4. Discussion

In order to parcel larger brain regions into functionally specific sub-regions we propose to use intrinsic functional dynamics present in BOLD FMRI recordings. To this end we apply independent component analysis to FMRI data that were transformed to amplify minor temporal differences between individual voxels within the larger brain region. Results of our strategy obtained for sensory-motor cortex and subcortex, as well as thalamus and enthorinal cortex showed that the boundaries of the ensuing, group-level, parcellations closely corresponded to the topographic organization observed in available cytoarchitectonic atlases and that the obtained subdivisions reflect biologically known areas of the brain.

### 4.1. Sensory-Motor ICP

The motor cortex has been the focus of a large number of studies. Early work by Penfield already demonstrated that there is a somatotopic organisation within the motor cortex, identifying separate regions controlling different parts of the body (Penfield & Boldrey, 1937). More recent work, both in non-human primates and in humans, has furthered our understanding of this somatotopic mapping, and introduced the concept of the homunculus (Rao et al., 1995; Schott, 1993), forming a detailed functional topography within the motor cortex that has previously been demonstrated to be identifiable in R-FMRI (M. P. van den Heuvel & Hulshoff Pol, 2010). Here, we were able to retrieve a parcellation of the sensory-motor system that followed the organization of the homunculus along the central sulcus, while at the same time also adhering to an anterior-posterior organization as present in the Jülich histological atlas.

It should be noted that the alternative parcellations, i.e., ICA and Craddock (scale 50) also showed considerable overlap with the Jülich Atlas, with moderate to high Dice's overlap comparable to the results obtained for ICP. This is in contrast to the results obtained for other regions where ICP showed considerably larger overlap with the atlas regions. We believe this difference is due to the generally larger functional regions present in motor cortex that are accordingly easier to pick-up on by different parcellation techniques. That said, when visually comparing our ICP results to the multi-modal parcellation obtained by Glasser et al., it is clear that the Glasser parcellation did not retrieve the medial-lateral homunculus organization (Glasser et al., 2016).

### 4.2. Subcortical structures ICP

The subcortical areas, in contrast to the motor strip, are situated much deeper in the brain. This inherently makes them more challenging to parcellate, as the sensitivity profiles of modern 32-channel MRI head coils distinctly favour the surface of the brain, leading to lower SNR in regions deeper in the brain. This is further complicated by the small volume and high functional differentiation of the subcortical areas. However, despite their small size, they form key nodes in the human connectome. Here we were able to accurately recover both coarse (8 sub-regions) and fine-grained (27 sub-regions) parcellations of the subcortical areas. The fine-grained parcellation showed a high degree of overlap with the known sub-regions in subcortex. This degree of overlap was not achieved by ICA without temporal unfolding or in the Craddock parcellation. Interestingly, the two levels of reproducible IC-parcellation accurately reflected two levels of granularity in subcortex. At the coarser level, we observe larger biologically meaningful substructures (e.g. pons, medulla, midbrain, and hippocampal formation), which further subdivide into smaller regions (e.g. hippocampus proper and amygdala).

### 4.3. Thalamus ICP

Previous studies have demonstrated how thalamus has a wide range of connections to all of cortex (Behrens et al., 2003; Johansen-Berg et al., 2005), with thalamus acting as a major functional hub, routing information between various parts of the brain (Achard, Salvador, Whitcher, Suckling, & Bullmore, 2006; Gili et al., 2013). Internally the thalamus is divided into several cytoarchitectonically identifiable nuclei, each with their own distinct patterns of connectivity (Behrens et al., 2003; Johansen-Berg et al., 2005; Morel et al., 1997; Zhang et al., 2008).Using FMRI data we were able to retrieve a plausible functional subdivision of thalamus. For example, we were able to accurately identify important nuclei including LGN and subdivisions of the pulvinar region. Similar to the results for the subcortical area's, the IC-parcellation was found to have a higher overlap with known histology (in the form of the Morel atlas) compared to a parcellation obtained using standard ICA or NCUT clustering.

### 4.4. Entorhinal ICP

Compared to thalamus, entorhinal cortex is a more specialized structure. Studies on rodents and non-human primates suggest that the EC projects to a wide range of structures, with portions of the medial EC projecting to the hippocampus, thalamus (specifically dorsolateral and dorsocaudal thalamic nuclei), amygdala, olfactory structures, visual, parietal, cingulate and retrosplenial cortices; while lateral EC portions connect to the septal hippocampus, basal ganglia, claustrum, amygdala, thalamus, as well as frontal, insular, and cingulate cortices (Canto et al., 2008; Kristin M. Kerr, Agster, Sharon C. Furtak, & Burwell, 2007; O'Reilly, Gulden Dahl, Ulsaker Kruge, & Witter, 2013). While the organization of EC is well known in the rodent, there is an ongoing debate as to how this translates to the human EC. Using the potential of ICP to obtain biologically plausible parcellations in small structures, our results provide further evidence for the inferior-superior and lateral-medial organization. It thereby replicates earlier work, where such an organisation was established on the basis of connectivity gradients (Navarro Schröder et al., 2015). Importantly, a fine-grained parcellation of the EC obtained non-invasively and in-vivo opens the door to study EC involvement in cognitive processing or their vulnerability to pathophysiology (Braak & Braak, 1992).

### 4.5. Alternative Parcellation Strategies

When comparing currently available parcellation strategies we propose to look at several key criteria that describe what a good parcellation strategy should adhere to in the context of systems neuroscience research. First and foremost, subdivisions should reflect underlying neuro-biological principles and should closely follow underlying functional segregations (Blumensath et al., 2013). Inaccurate parcels spanning multiple underlying functional regions significantly reduce performance in the estimation of network topology (Smith et al., 2011). Second, the parcellation needs to be able to produce viable results across the entire brain, to prevent incomplete network representations. Third, the obtained parcellation should allow group-level analyses, assessment of group differences, or the generation of a parcellation that is specifically tailored to a group of interest. Finally, to ensure usability and user friendliness the parcellation strategy should be fully automated.

While existing parcellation strategies (e.g., random parcellation (Hagmann et al., 2008), NCUT (Craddock et al., 2012; M. van den Heuvel et al., 2008), or region growing (Bellec et al., 2006; Bellec, Rosa-Neto, Lyttelton, Benali, & Evans, 2010; Blumensath et al., 2013)) include extensive automation and can be applied to the entire brain, they typically do not satisfy the other criteria of an ideal parcellation. A random parcellation for example, generates ROIs using an algorithm based on a random process. Subsequently, network measures are calculated over several iterations of the parcellation to obtain a robust set of metrics. This makes it highly likely that each parcel contains contributions of several underlying functional segregations. Accordingly, random parcellations do not follow underlying functional divisions, nor do they employ biological information. In contrast, region growing, and to a lesser extent NCUT methods, employ various clustering algorithms to parcel the brain, exploiting underlying functional information. It is possible for region growing algorithms to intelligently specify initial seeds reflecting underlying anatomical and/or functional brain organizational principles. To obtain parcellations that are independent of brain state, these algorithms are commonly used on metrics derived from unconstrained resting state FMRI scans. As an example, NCUT spectral clustering can be used to create clusters by maximizing internal similarity while minimizing the similarity between clusters (Craddock et al., 2012; M. van den Heuvel et al., 2008). In contrast, region growing approaches start from initial seeds to subsequently create a hierarchical clustering tree (Bellec et al., 2006, 2010; Blumensath et al., 2013). Common to these parcellation strategies is that they tend to generate clusters of uniform size (and to a lesser extent, shape (Glasser et al., 2016)). It is doubtful that this forms an accurate reflection of underlying biology (David C. Van Essen et al., 2012). This is also evident from the results presented here, showing that the Craddock NCUT-based parcellation did not overlap well with the known histological atlases.

Finally, it is often particularly challenging to generalize obtained parcellation schemes across participants to investigate group-level effects. As there is no underlying biological basis for a random parcellation, it cannot be generalised across participants to generate biologically meaningful results. This makes it for instance impossible to assess volumetric differences between a control and a patient group. NCUT and region growing techniques often operate at the single subject level. While bringing individual parcellation schemes to the group-level could be achieved through averaging or group-level clustering, such approaches typically fail to preserve the detailed parcellation obtained for each participant. This seems to be an inherent problem to the clustering methods, as (Yeo & Krienen, 2011) using 500 participants could at most distinguish 17 separate clusters. In ICP parcels are generated immediately at the group level (although ICP can also be applied to individual data). This has the benefit of generating results that are valid at the group level and that can be used to directly compare two groups, for instance with respect to the size of the obtained regions.

### 4.6. ICP considerations

We want to highlight the potential of ICP to obtain parcellations of structures that are hard to parcel due to their location in the brain or their very convoluted folding patterns. As an example the recent multi-modal brain parcellation proposed by Glasser et al. (2016) concentrates on cortical parcellations. The ICP strategy presented here only relies on the availability of R-FMRI data and is thereby able to resolve sub-regions within smaller, deep brain structures, especially when incorporating high-resolution imaging.

We optimise for biological validity of the ICP through the use of split-half reproducibility. The use of ICA-denoised data is crucial to this step, as this will remove structured artefacts from the data. Under ideal conditions this leaves correspondences of underlying biology as the only signal source to modulate split-half reproducibility when a range of scales are considered. Split-half reproducibility does however have some caveats to consider. Most ICP split-half reproducibility results show a downward slope with increasing number of sub-regions. This bias is caused by the method used to calculate the match between parcellations. As the number of sub-regions increases, the size of these sub-regions decreases. Two large sub-regions which are shifted by one voxel with respect to each other will still show a high Dice overlap. Two small sub-regions shifted by one voxel will show a substantially lower Dice overlap. Final scale selection should therefore be based on local maxima in split-half reproducibility.

A second issue to be considered in the context of the split-half reproducibility in ICP is the occurrence of peaks at multiple scales. Each of these parcellations might be equally valid, and can represent splits at different levels of the hierarchy of the investigated structure, as can be seen in the subcortex results presented in this work. In this parcellation, the hippocampal formation is found as one single structure at a coarser level, where it splits up into the hippocampus proper and amygdala. Both the coarse and the fine-grained parcellation are biologically valid, their usefulness depending on the specific research question at hand. The underlying hierarchy together with data quality and quantity will determine up to what level of detail parcellations can be reliably estimated. SNR will be a limiting factor in finding reproducible parcellations in a given dataset. Here we maximize SNR by using a large, high quality dataset in the form of the Human Connectome Project.

### 4.7. Conclusion

We have introduced a novel approach for parcellating the brain in a top-down fashion on the basis of instantaneous connectivity between voxels of a region and the region-specific average temporal dynamics in a top-down fashion. Our technique allows for the gradual refinement of the ensuing characterisation of the region and its sub-components. By exploiting small differences in the temporal dynamics of sub-regions within larger ROIs the ICP strategy is able to retrieve biologically plausible parcellations. Not pinpointing one-self to a specific scale of parcellation, ICP will allow researchers to investigate ROIs in a hierarchical manner, even in areas of the brain that can only be resolved with high-resolution imaging. For future work we will use ICP to generate a hierarchical, full brain parcellation.

## Acknowledgements

Data were provided [in part] by the Human Connectome Project, WU-Minn Consortium (Principal Investigators: David Van Essen and Kamil Ugurbil; 1U54MH091657) funded by the 16 NIH Institutes and Centers that support the NIH Blueprint for Neuroscience Research; and by the McDonnell Center for Systems Neuroscience at Washington University. C.F.B. gratefully acknowledges funding from the Wellcome Trust UK Strategic Award [098369/Z/12/Z]. C.F.B. is supported by the Netherlands Organisation for Scientific Research (NWO-Vidi 864-12-003). M.M. received support from a Marie Curie International Incoming Fellowship from the European Research Council under the European Union's Seventh Framework Programme (FP7/2007-2013) / ERC grant agreement n° 327340 (BRAIN FINGERPRINT).

# Supplementary materials for *Functional parcellation using time courses of instantaneous connectivity*

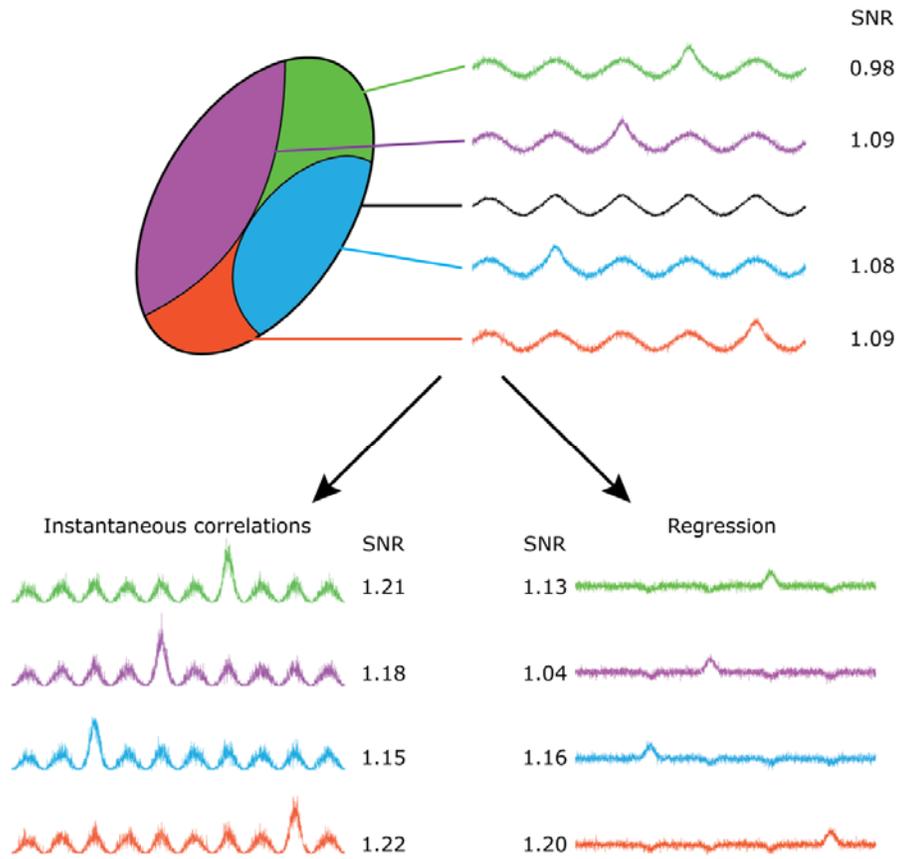

**Supplementary figure 1:** ICP compared against regression of the mean ROI signal. While the increase in SNR between ICP and regression is similar, regression results in a greatly reduced variance in this example.

| 101309 | 154936 | 198855 | 433839 | 814649 | 102311 | 128127 | 155635 | 173940 | 201818 |
| --- | --- | --- | --- | --- | --- | --- | --- | --- | --- |
| 103111 | 164131 | 203418 | 445543 | 833148 | 105014 | 131217 | 157437 | 177645 | 205220 |

| 116524 | 171633 | 208327 | 573249 | 837560 | 106521 | 131722 | 159441 | 178142 | 209834 |
| --- | --- | --- | --- | --- | --- | --- | --- | --- | --- |
| 122620 | 172029 | 211215 | 594156 | 871762 | 108121 | 132118 | 160830 | 179346 | 210415 |
| 142626 | 172130 | 212116 | 599671 | 901038 | 108323 | 133019 | 161630 | 180129 | 211922 |
| 147030 | 173334 | 308331 | 623844 | 910241 | 111413 | 135528 | 164939 | 181232 | 212419 |
| 148941 | 187143 | 352132 | 690152 | 912447 | 118528 | 140117 | 167036 | 189349 | 231928 |
| 149741 | 191336 | 380036 | 695768 | 922854 | 120515 | 141826 | 169444 | 191033 | 290136 |
| 150625 | 192843 | 385450 | 742549 | 958976 | 123420 | 145834 | 171431 | 191841 | 303119 |
| 152831 | 194645 | 395958 | 789373 | 983773 | 123925 | 146331 | 173435 | 197348 | 316633 |

**Supplementary table 1:** Subject ID's of the data used for parcellation from the Human Connectome Project.

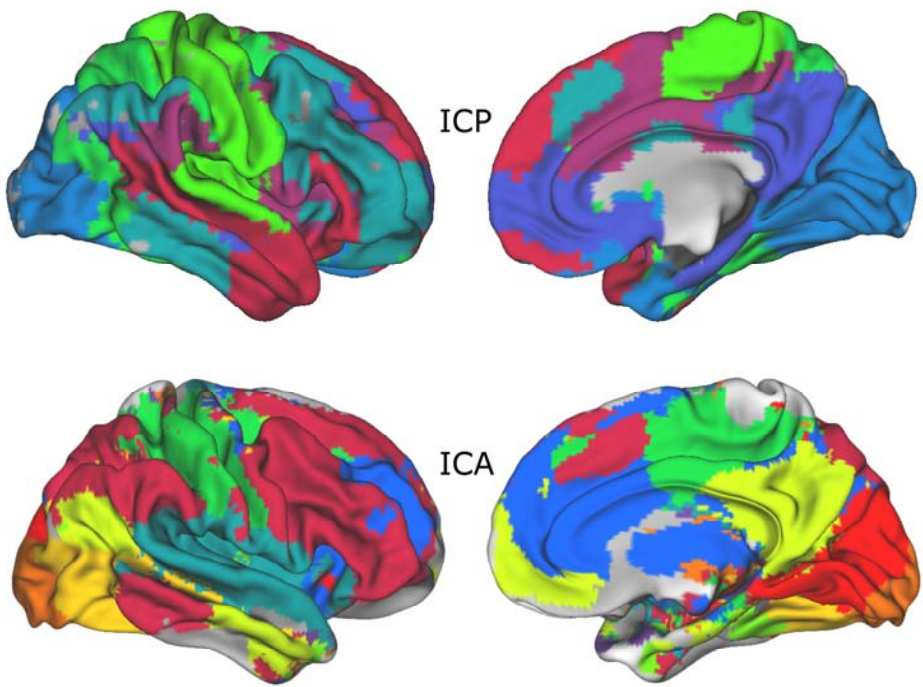

**Supplementary figure 2:** Full-brain ICP (top row) compared with ICA (bottom row) by Smith et al. (2009). The sensory-motor system was selected from this ICP result and used for a next-level parcellation.

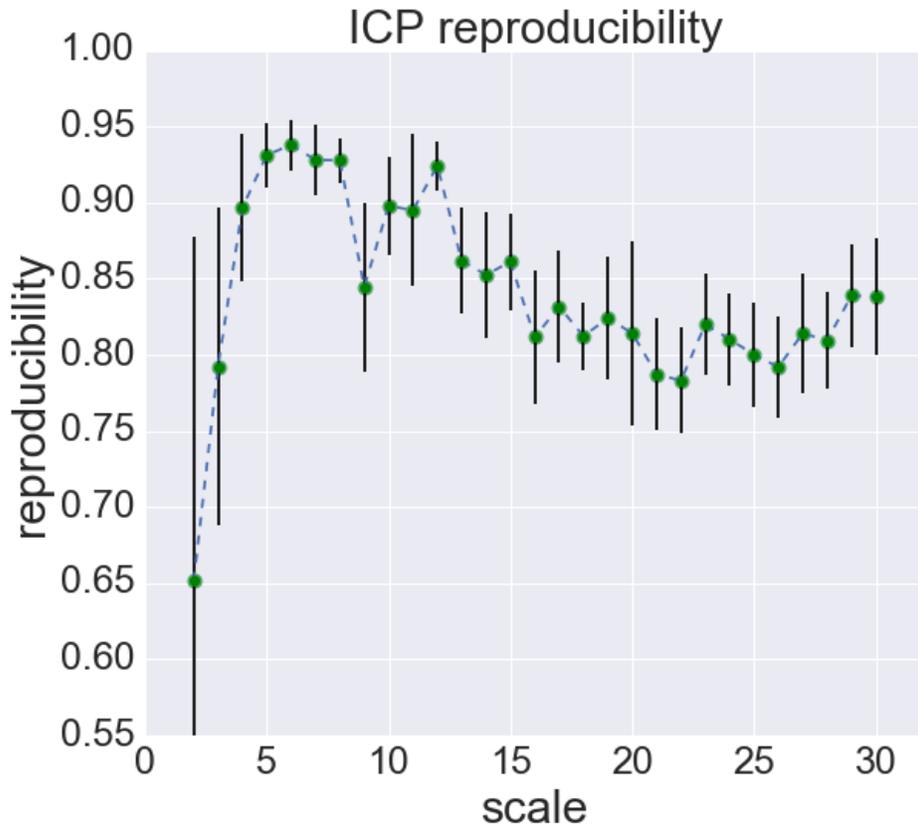

**Supplementary figure 3:** Split-half reproducibility of the sensory-motor cortex parcellation.

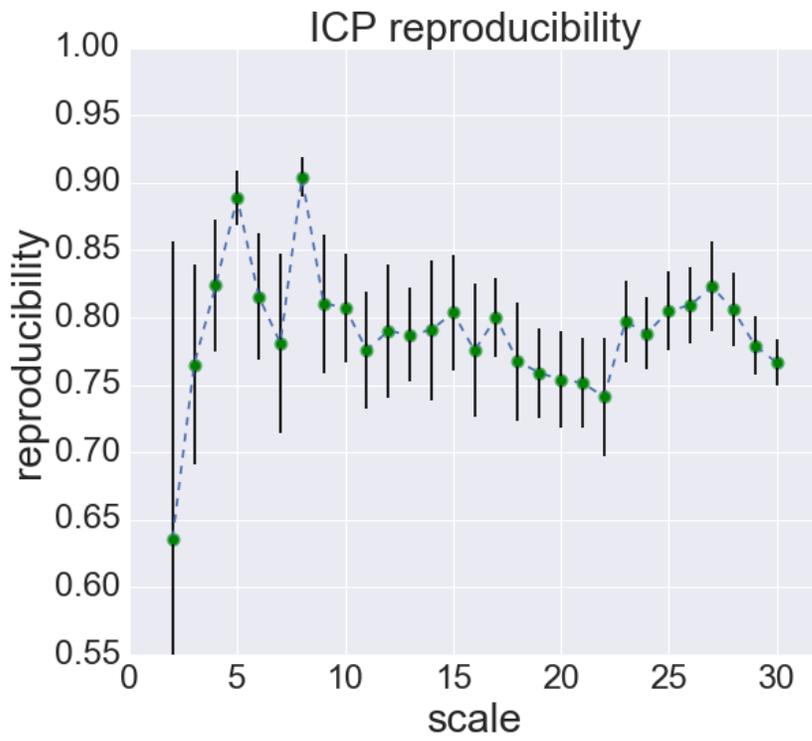

**Supplementary figure 4:** Split-half reproducibility of the subcortical parcellation.